# The Role of Resource Awareness in Medical Information System Life Cycle

Petar Rajković
Faculty of Electronic Engineering
University of Niš
Niš, Serbia
petar.rajkovic@elfak.ni.ac.rs

Anđelija Đorđević
Faculty of Electronic Engineering
University of Niš
Niš, Serbia
andjelija.djordjevic@elfak.ni.ac.rs

Aleksandar Milenković
Faculty of Electronic Engineering
University of Niš
Niš, Serbia
aleksandar.milenkovic@elfak.ni.ac.rs

Dragan Janković
Faculty of Electronic Engineering
University of Niš
Niš, Serbia
dragan.jankovic@elfak.ni.ac.rs

*Abstract*— During the process of medical information system development, resource awareness is neglected. It is often assumed that the underlying hardware will always have enough memory, processing power, and network bandwidth. Unfortunately, this approach seems not so feasible in every case, and these assumptions, if proven wrong, will harm the initial development run, a later system upgrade, and life cycle in general. This paper aims to raise a resource awareness problem that could still influence all information system deployment and maintenance steps. As an example, we described the influence of the general hardware and network limitations on the information system design, functionality update process, and external system integration. Our research results are a set of guidelines that should be applied to support resource awareness, as an important part of the system design.

## I. INTRODUCTION

The development of information systems becomes the process where resource awareness is neglected [1]. It is often assumed that the underlying hardware will always have enough memory, processing power, and network bandwidth [2]. Nevertheless, this opinion does not come out anywhere, but from the fact that the infrastructure has become increasingly reliable in the past decades. Besides that, mission-critical systems, such as Medical Information Systems (MIS) certainly are, must be designed with the appropriate level of redundancy. The end-user software must be operational, regardless of the status of the underlying network, or server infrastructure.

Unfortunately, many modern-day MIS systems are designed in a way that their process execution often depends on the resources that are outside of their control and the possibility to influence. The approach that includes heavy dependence on the external systems seems not so feasible in every case. However, if these assumptions are proven wrong, they will have a negative effect on the initial development runs, later system upgrades, and life cycle in general [3].

In ambulatory and primary medical care institutions, medical information systems are in use in the Republic of Serbia for longer than a decade and a half [4]. The MIS developed by our research group, named Medis.NET, is active since 2009 in 25 institutions in southern and eastern Serbia [5]. It is intended to support daily jobs both for General Practitioners (GP), pediatricians, laboratory, radiology, and many different specialistic services. During this period, we were facing different challenges while developing the updates that aim to introduce new functionalities [6] [7], integrate with external systems, and improve the system's effectiveness [8].

The upgrades and improvements mentioned often required the data exchange with the remote services hosted in the area outside of the responsibility of MIS itself. Since any of these connections could be interrupted or even disconnected due to the various connection or maintenance problems, our MIS had to be designed in a way to be immune to such interruptions. Also, the internal system updates could leave the MIS clients in detached mode, discontinuing the medical professionals from their regular daily business.

In the following sections, we place an overview of related work both from medical information system development and general software design methods. Related work is followed by the description of the system architecture and external system integration scenarios. In the discussion section, we pointed out the influence of the resource awareness approach on the presented solutions.

In this position paper, we present the challenges we had during the system life cycle with an emphasis on the requirements that come because of the resource shortage and the necessary management procedures that we had to develop and apply. The described procedures are summarized as the set of general observations that could apply during separate phases of the medical information system lifecycle.

## II. RELATED WORK

Since this paper tends to summarize our activities in the MIS life cycle maintenance, the set of related work which was in our focus could be divided into the following topics:

- Technology stack choice
- MIS architecture choice
- System upgrade approaches

The positive point in technology acceptance is that from the user's point of view, the more crucial factor is the system effectiveness that technology used [9]. Regardless if the system is designed for telemedicine or Electronic Health Records (EHR) the quality of the solution is more important [11] [12] [13].

Since this sentiment [14] was not significantly changed over time, we decided to base the choice of the technology stack on the general technology acceptance and resource availability. The next most key factor was the development team experience [15]. Thus, we chose Microsoft's platforms since they offer the full technology stack [16] starting from the database, via Service-Oriented Architecture (SOA), Web, and Windows-based clients up to cloud infrastructure [17]. Also, all of these are supported by state-of-the-art integrated development tools [18], community portals, and helpful resources [19].

When it comes to MIS architecture choice, we chose an extendable SOA-based approach [20]. Compared to the basic two-tier and three-tier application models, SOA gives a better flexibility level and made all system extensions, that would be required later, much easier [21]. The only choice that brought a slight flavor to the two-tier approach is the decision to provide the thick client for the medical professionals [22]. This decision is considered valid even for the different architecture approaches which move significant parts of storage and data processing to the cloud. Furthermore, the work presented in [23] suggests the same solution proposed for different fog networks. Besides the other approaches examined (such as thin and Web clients), we carefully decided in favor of the thick clients having in mind all the possible benefits in the cases when underlying resources become less reliable [24].

Regarding the upgrade strategies, we decided to go for these that will help us in reducing the application downtime during the system update. Looking in this direction we have chosen blue-green deployment for the one-point running services [25] and canary deployment for the applications installed on multiple instances [26]. This is a common choice and is elaborated on in detail in multiple diverse types of research. The approaches such as dark mode and feature flag deployment [27], are not included since they would require much more effort for the development [28], while the effect will not be in the expected equivalent.

## III. SYSTEM DESIGN CONCEPTS

Before we start to develop the initial version, we had to choose the technology stack, the system architecture, the update approach, and the extension points. At that time, the requirements were drawn by the Serbian Ministry of Health with the focus only on the set of basic business requests and the functionalities that would be required only by GP in primary health centers (Fig. 1).

The infrastructure in the target medical institutions was underfunded, and, in most cases, not up to date. For reference, the area where our MIS is installed is the part of the country with the lowest population density and is economically less developed than the country's average.

There were no plans, nor mentions of any potential extensions, and without any further upgrade plans. Since it was obvious that at one-point systems should become extendable and with data exchange possibilities, we decided to go towards the service-oriented architecture to its full extent, but the infrastructural limitations we saw in our target institutions, made us modify the approach from the beginning.

### A. Hardware Limitations

As was mentioned before, in many cases, the public sector, including the health infrastructure, is underfunded. This results in outdated hardware, general applicative software, and operating systems. Unfortunately, but understandably, the IT infrastructure usually comes after medical and general equipment when medical institution renovation projects get launched.

To illustrate the previous claim, we could mention that client health institutions were able to provide only one server that should play the role of both database and application server. In some cases, the same computer acts as a domain server too. Next, client machines are usually the cheapest configurations that can be found on the market, at the given moment. The client computer in the doctor's office is used as a general-purpose computer and not only as an MIS dedicated machine.

The network installed in the institution usually meets the standard that was now when installed. Once installed the network will remain the same for a prolonged period. I.e., we have no institution that has significantly updated its network infrastructure after the initial installation. Nevertheless, internal networks are at an acceptable level, but when comes to communication outside of the institution, connections could be slow and unstable.

### B. Basic System Architecture

Having in mind all the technical constraints mentioned, we had to choose between a Web-based solution, fully service-oriented architecture with thin clients, and partial SOA relying on the database server and thick clients.

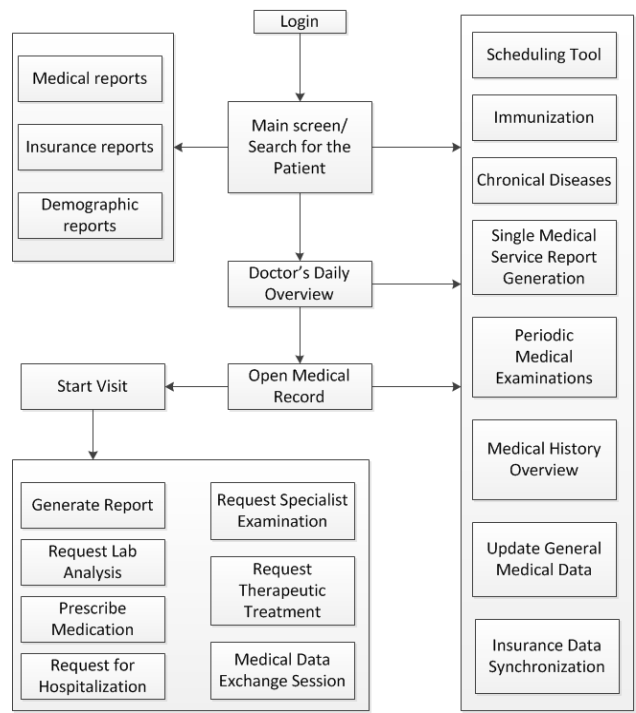

Fig. 1 MIS - common functionalities scheme (as in [7])

The web-based solution looks like the easiest from the end-user's point of view. There is no need for the additional piece of software, and it could be run through the Web browser. Unfortunately, they depend heavily on the network and any problem in the network will immediately reflect the end-user satisfaction. Next, the technology stack will require additional client-side technology such as different script-language-based frameworks.

In choosing the client application, the technology stack could be reduced to some framework that supports many different Windows and SOA applications, such as Microsoft.NET. The abovementioned developing ecosystem was our first choice due to the, at that time, best support network.

Eventually, we choose thick client architecture since it proves to be the optimal solution for the environment where it should run.

A thick client is a larger application than a thin client but requires no constant active connection with the server to run. This is especially important where the network is considered as a potential issue. The thick client could accumulate data and synchronize when possible. With the thick client, the users will always have available all the major functionalities and only a few of them, which require a permanent network connection, will be disabled.

Next, we choose to develop both server and client-side software as modular and plugin-based. This allows the parallel development and independent update of the components that should be changed or added during the software life cycle.

The technology stack that we chose consisted of Microsoft.NET as the general framework where all the applications were developed, supported by NHibernate object-relations model and PostgreSQL database.

*C. Functionality Updates*

The software update is based on two major strategies – blue-green deployment for the service components and canary deployment for the clients.

When performing the blue-green approach in Fig. 2 We have minimal downtime for the service components which is then reflected in the overall system readiness. Even in the case when the deployment fails and the rollback to the previous version is fast, and together with the thick client approach, the system suffers from downtime as little as possible.

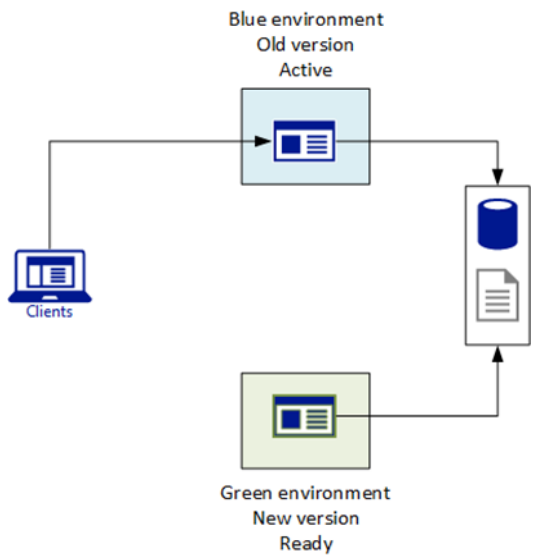

Fig. 2 Blue-green deployment scheme

The update of the clients is done through the canary deployment approach (Fig. 3). When the updated version is developed, it is pushed to the selected set of clients. The selected clients are chosen because of representatives – in the sense of functionality usage and data frequency. Once deployment is proven in canary clients, it is pushed forward in the complete network.

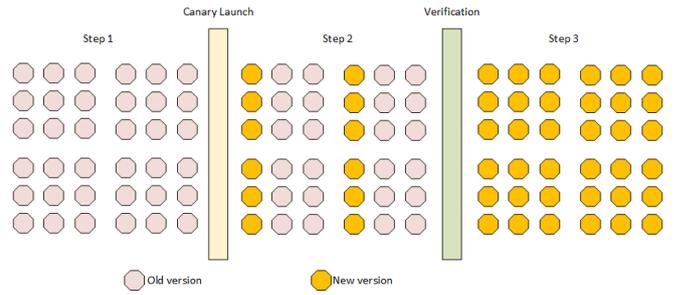

Fig. 3 Canary deployment

The modular-based client is one additional point that helps in general resource awareness. Since the client is configured as the set of plugins, only these clients that contain updated components will be targeted. Since any software upgrade requires package distribution through the network, any reduction is in favor of resource awareness.

The additional benefit of the modular-based software is that the update packages could consist only of the changed component. Since many times, only limited bug fixes and updates must be applied, a lighter package ensures more efficient distribution and fewer potential problems due to network usage.

IV. EXTERNAL SYSTEM INTEGRATION

Medical information systems were initially designed to cover the needs of one healthcare facility. Therefore, all patient data were kept locally and could be accessed only from the Healthcare Institution (HI) that has created them. Over time, the need for different data integration was born (Fig. 4).

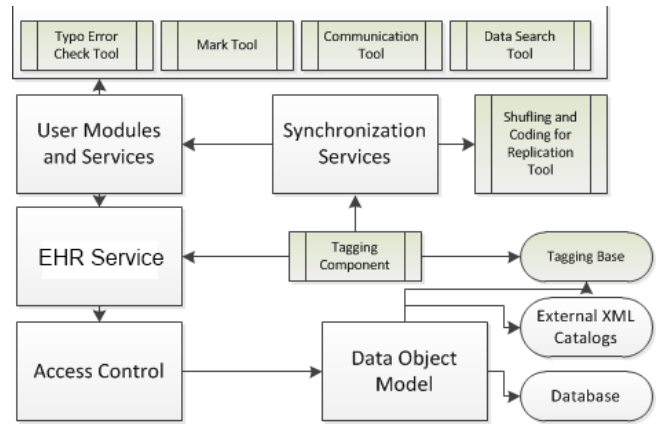

Fig. 4 Basic data integration model (as in [10])

*A. Integration with Insurance Related Services*

First information systems for healthcare facilities were developed for recording material costs and creating invoices, therefore integration with insurance-related services represents an important part of MIS. Republic Health Insurance Fund (RHIF) created the report format that aims justification of spent materials and medicines and keeps records of provided health services (Fig. 5 and Fig. 6). HI should usually submit the report monthly. The creation and data manipulation of the report is developed in a module in Medis.NET [29]. The report format is overly complex and contains a lot of data details. For the report to be accepted, it must be formatted correctly, respecting predefined rules, and must contain valid data.

The created module for invoicing the provided health services collects data, and processes, and exports them to an XML file (electronic invoice), which is the required form by the RHIF. The XML report is created on provided health services but written with the RHIF codebook. The module does not need additional hardware and does not cause an additional financial cost for a HI. However, the RHIF codebook is frequently changed, therefore the module must keep up with those changes. The codebook is in XML format, as well. A new codebook version is imported into the module via the XML2SQL tool, which is specially developed to import XML data into a relational database [30].

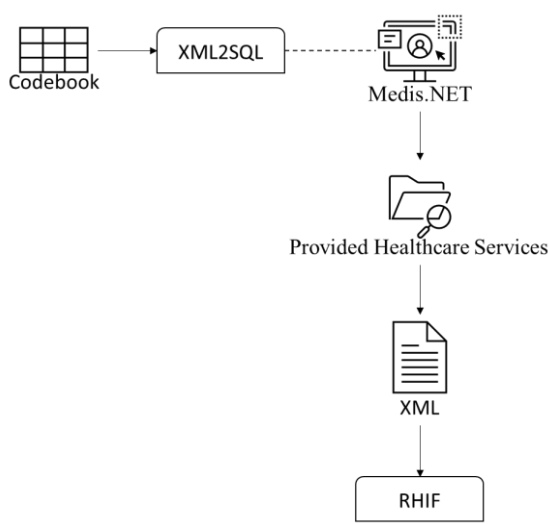

Fig. 5 The integration of Medis.NET with insurance-related services

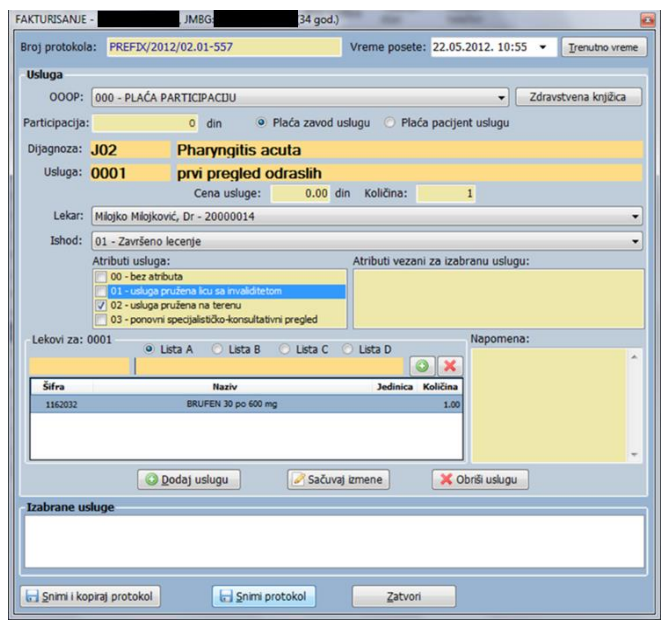

Fig. 6 Integration with the insurance system

### B. Integration with Medication Management Software

Every HI on monthly invoices needs a list of used medications inside the institution. Name similarities between different medications can cause a big problem while invoicing the provided healthcare services. Name differences can reflect in medication strength, shape, number of doses per package, etc., and still, it can be the same medication. Also, the name of one medication may vary depending on the pharmaceutical manufacturer. Slight differences between medications can easily cause errors on a daily level, which will be visible when RHIF receives an electronic invoice. RHIF will reject all invoiced healthcare services with medications that HI did not have in stock or which price was different from RHIF's determinate price in that period. After the rejection, HI needs to create a new invoice for the same month and send it again to RHIF. For all the reasons above, the need for the usage of medication management software was born. During that time, many applications with similar purposes were used in Serbia and wider, but HIs in southern and eastern Serbia did not have enough financial resources to afford them. Another problem and the additional cost would represent their integration with our MIS system. Therefore, the software module for medication management was created and integrated with Medis.NET [31].

The software module provides medication tracking in HI. It allows invoicing only for medications that are in storage in a central warehouse or warehouses in different departments of HI, where the healthcare service is delivered. The module allows the ordering and issuing of medications and creates an alert when the amount of medication is below the minimum. There is also an alert for medication whose expiration date is near. The module provides medication transfer between a central warehouse and local warehouses inside healthcare organizations. The created module can work as an individual application or in integration with MIS Medis.NET.

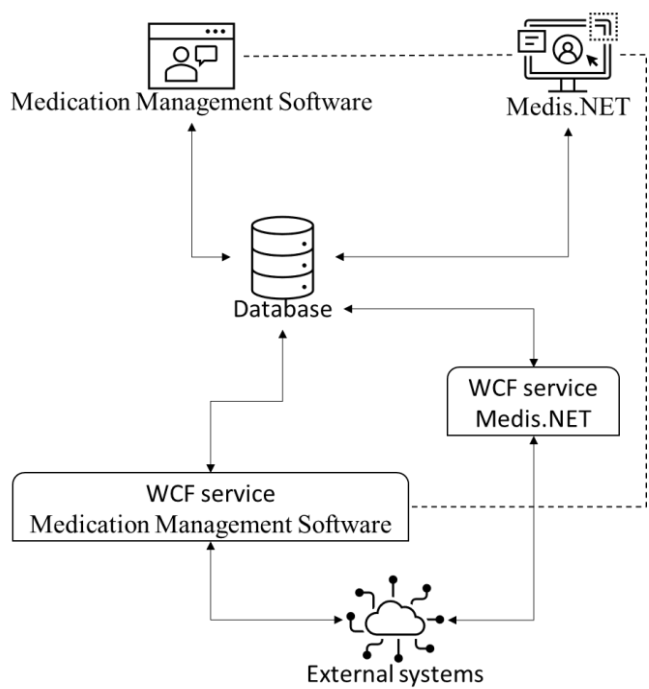

Fig. 7 The integration of Medis.NET with medications management software

The integration was realized via web methods available over Windows Communication Foundation (WCF) service or directly, using the appropriate libraries. WCF is Microsoft's framework for building service-oriented applications [32]. WCF service was chosen for the integration because it is a part of Microsoft's technology stack, which our system uses. This service provides module integration with other external systems as well, therefore this way of realization allows the system to be open to other software on the market. The integration, on the other hand, disables invoicing of medications out of stock, or with wrong prices on Medis.NET. The software module developed in this way gave the best solution for preventing medication invoicing errors in HIs with limited financing resources. The integration of Medis.NET with medications management software is shown in Fig. 7

*C. Integration with other MIS Systems*

Another case of Medis.NET integration with external systems is patient data access through different MIS systems. Keeping patient data locally and providing access to patient data only from the healthcare institution which created them disables tracking patient's previous medical treatments while visiting other healthcare facilities. Therefore, the partial collaboration between heterogeneous MIS is established through the exchange of data on patient visits, referrals, and reports [33].

The first direction of communication is established only when a current patient has had previous treatments, which are important for the current visit to another HI. MIS sends a request to the Central Repository (CR) and waits for patient data in response. This communication direction does not need any extra resources but requires an Internet connection. Since this communication is established only for patients with treatments in other HI, and only when it is needed, which is not very often, this approach will not take a lot of network capacity in the average case.

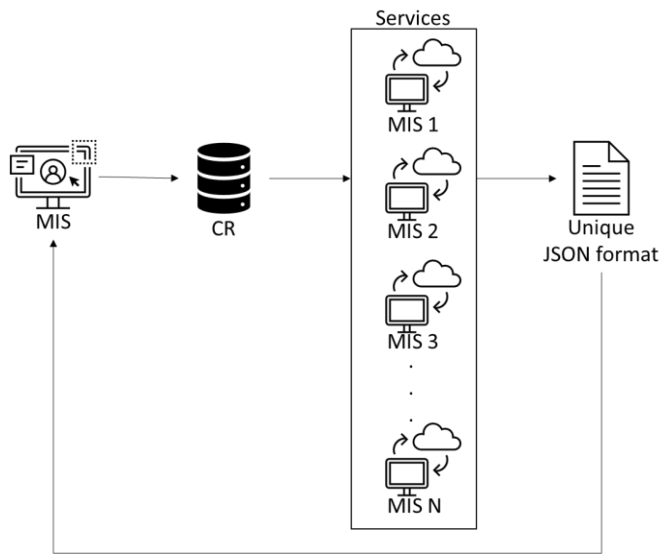

Fig. 8 The collaboration of heterogenous MIS systems

The second direction of communication is established when MIS contains patient data that some other MIS needs. In that case, MIS receives requests for patient data, collects, packs, and sends the data back. For this communication part, each HI needs one server with an installed service that receives patient data requests and gives a response. This communication direction does not impact normal client work but requires installed service on a server machine. The collaboration of heterogenous MIS systems is shown in Fig. 8.

*D. Integration with other "in-house" healthcare information systems*

Besides accessing data on patient visits, referrals, and reports it is important to make all radiological recordings and findings available to all MIS, regardless of where the recording was performed. Therefore, the Central Radiological Information System (CRIS) was created [34]. Each MIS sends a radiological referral to CRIS and can access the radiological report and radiological image after the examination. In Medis.NET two modules for collaboration with RIS are created, and HI can choose which one will use (Fig. 9).

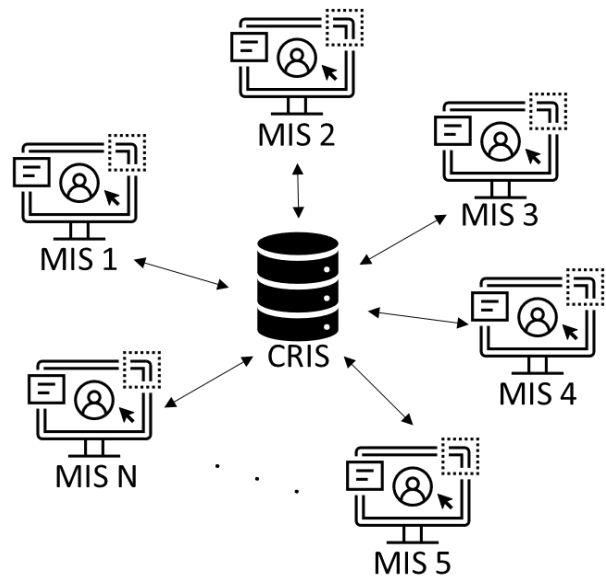

Fig. 9 Integration with other "in-house" healthcare information systems

"On-demand module" is used when a patient comes to the physician who created the referral, after the radiological examination. MIS sends a request to CRIS to obtain the examination results. This module does not require any additional investment by HI, and will not slow down the client's work, but there is a problem when a patient does not visit a physician after the radiological examination. Therefore, services provided in the HI will not be invoiced, and will not be charged to the RHIF.

"Webhook module" enables automatic data transfer on the realized radiological examination to Medis.NET. When a patient data change is made in CRIS, a notification is sent to the MIS of HI that created the referral. All changes are automatically recorded in MIS. This module does not have an invoicing problem like the previous one. However, this module requires a special server for its implementation, as well as a static IP address. Implementing this module will cause additional financial investments for HI.

## V. DISCUSSION

The life cycle of one software product, especially one custom-built for the dedicated customer, goes through many points where updates are required, and even changes in the technology stack must be applied. Our clients have actively used Medis.NET for more than twelve years and helped the system improve through the different suggestions and change requests.

These requests often brought a significant level of complexity into the MIS, but the infrastructural updates rarely follow. For these reasons, we had to carefully design every aspect of the system and sometimes made changes even in the technology stack.

*A. Technology Stack Changes*

General feature addition, bug fixes, and usability enhancements are regular updates, expected for any software. Mentioned changes in the technology stack are not so often the case. It came usually after significant market or infrastructural change. It is important to note that this kind of change could result in many hours of unintended work in development and testing. Looking from the database perspective, our first choice was Microsoft SQL Server since we rely on Microsoft technology.

Microsoft SQL Server has many different versions, and it is one of the most powerful relational database engines nowadays. Unfortunately, it is not license-free software. The mentioned funding problem soon put us in the situation where we should replace the chosen database server with a new one that is open source, and its usage is free of charge.

Thus, we switched to PostgreSQL. Besides, we used ORM and did not have direct access to the database, some problems had to be addressed directly. I.e., any report query that contained a "LIKE" statement had to be updated with "ILIKE" due to the SQL dialect differences. Apart from this, the other differences were considered minor and for a period of a week, we could finish all the connected tasks. PostgreSQL was powerful enough to replace SQL Server for the clients such as medical centers where our software was installed. The most demanding deployment has less than 500 simultaneously active clients, thus our database replacement was proven adequate.

The next significant challenge was a change of ORM. We started with Microsoft's Entity Framework (EF), and later switched to NHibernate. The main reason for this change is the performance during the loading phase. Earlier versions of EF had disadvantages during the start phase, while during the execution there is no significant difference between EF and NHibernate. The reason was the fact that the EF requires the creation of data structures representing the database schema, which takes a lot of unnecessary space and time. The change was to define an abstract data layer, which would allow us to create necessary classes, by the mode of the adapter and abstract factory pattern, that will allow any future ORM model changes. The approach was also successfully used when CQRS-based reporting was introduced [35].

*B. System Architecture*

When looking at the landscape of contemporary solutions, the thick client approach is still viable and widely used. The same approach is used by the well-known NextGen Office [36]. The software mentioned also supports integration with external systems and data exchange, as well as Medis.NET. On the other hand, solutions based on thin and Web clients are more common solutions, especially for the cases where the stable connection is considered the constant. The most used solution is Serbian primary care, the Heliant [37], which is Web and cloud-based as well as Centricity EMR [38].

The thick client approach in our case is still the preferred way to continue since it proved to be a solution that, in regular use, consumes the least of the network capacity since only data is transported over the network during the software's uptime. The problem that is important to mention here is data caching and data synchronization. To handle these requirements, we choose to use the local data storage which is periodically synchronized with the central database.

In metropolitan areas, primary medical institutions often consist of a few central larger and many smaller facilities. The number of active clients could vary between a few stations and up to several hundred. When the network connection is stable, the clients access the central database and update the local data storage in the background. When the network connection becomes slower or less reliable, thick clients switch to local storage as their primary data source.

To handle the mentioned data replication scenario in distributed data environment we have developed a platform-independent database replication solution, that can enable fast replication even in low-band/low-speed internet connections [39].

*C. Software Update Routine*

The challenge with the updates in ambulatory medical centers is that they work at full capacity in two shifts, while during the night, only the emergency remains active. The risk of erroneous updates is increased with the fact that the system must remain in high readiness constantly. The downtime caused by the update itself plus the rollback time would cause patients to queue and the quality of the medical service would suffer.

The update strategy must be then applied in a way that the regular operations remain as stable as possible. Thus, we decided to go for a blue-green deployment strategy for all service-based applications. The downtime will be only during the version switch, and in meantime, clients will remain functional due to the thick client architecture. Clients could provide the basic functionalities and the complete process will continue when the services come back online.

Client updates will be executed according to the scheme in the previous section. This will allow the end-user to continue her/his work until the updated version is downloaded. When the user finishes the current task, the only required action would be a client restart.

*D. Overall, Resource Usage*

The Medis.NET itself consists of many different applications and services that run as a single large system. The volume of the client depends on the set of configured modules and plugins and could vary significantly in terms of the space needed on the disk and space in memory while the application is running. The standard module composition needed for basic GP's work needs about 30MB of storage space on disk and around 120 MB in operational memory.

When it comes to the module that supports various specialistic services, the amount of needed space on a disk rises to 200 MB. The amount of memory needed for the imaging and radiological services is close to a gigabyte when processing large video files.

The amount of daily generated data per client is also dependent on the set of configured modules. While GP's clients have daily flow at the level of a few dozen megabytes, loading only a single video obtained by an ultra-sound device would need close to half of the gigabyte. Newer state-of-the-art devices, such are 3D scanners, generate a few gigabytes per session.

Next, each integration service runs as at least one separate Web service on one software node. The amounts of data they need are in the range of 100 MB, for prescription service, and up to 700 MB for the radiology exchange service. The traffic they generate per exchange session (at least one per medical examination/treatment) is in the range of 100 kB per a single recipe, via several MB for medical images up to gigabytes for the video.

One central database is running in the main database server, while the local data nodes are distributed, and their number could vary and depends on the spatial distribution of the customer's organizational units. The volume of the central databases is in dozens of gigabytes, while the local ones in the distant facilities are usually under 5 GB.

Different synchronization services run daily. They are responsible for exporting and importing different reports, updating catalog values, and ensuring data synchronization.

All the mentioned elements also require close monitoring, constant development, and support. For this reason, in the next section, we will not expose only guidelines related to a technical solution, but also to programming and technical and organizational efforts.

*E. Resource Awareness Guidelines*

To summarize our development efforts, we could say that all resource awareness actions during the software lifecycle could be divided into three groups, dedicated to reducing:

- data traffic and usage
- programming efforts
- downtime during the system update

**Use a layer of abstraction over the database and ORM**. Besides the change for these two parts of the system will not be frequent, it could happen from time to time during different circumstances – such as performance updates in one of the solutions from the market or even changes in pricing and licensing policy. This approach will save the time needed for the updates and make the updated version ready faster.

**Use lazy loading whenever possible**. Since many objects needed during the program execution contain a lot of fields carrying pieces of information that are rarely used, it is considered more efficient to load, by default, only object unique id and an indicative descriptor. If more details are needed, data will be loaded only upon the user's action. For example, medication is described with 22 different properties, but when displaying data about the prescription it is most important to display the medication name and dosing scheme. The remaining data will be loaded only when the user needs to update or check.

**Use data caching and distributed datasets**. To reduce data traffic as well as to handle the unstable network connection we suggest the usage of data caching for all catalog sets which are rarely changed, and which changes could be synchronized periodically with a low-risk rate. Distributed datasets are a next-in-the-line suggestion since they will allow a single medical facility to run even in the case when the connection with the central database is lost. The risk of data loss, in this case, exists in the situation when one patient needs to visit multiple medical facilities during a single day. To reduce the risk of this problem, the patient will get a printed document with a QR code carrying enough information that the record of his medical treatment or examination could easily be restored in the destination institution.

**Envise common extension points and support as much as integration methods**. This approach leads to later standardization when it comes to the connection to the external systems. Therefore, standardization will reduce design and development phases and help in data traffic estimation.

**Choose a software update routine that will reduce potential downtime, but not on the large programming cost**. Approaches such as dark mode or feature flags are considered safer than blue-green and canary, but their implementation came with a high development cost. To implement them properly, a significantly higher number of programming hours is needed. The downtime level achieved with blue-green and canary is considered acceptable in our case.

VI. CONCLUSION

Besides computers and various portable devices becoming highly capable, resource awareness is still a crucial point in information system design and development. The equipment itself could solve some scalability problems, but the proper approach in every step of the software's lifecycle could give a much better effect. In this manner, it would be significantly easier to conduct and maintain all the tasks and procedures related to the MIS.

Starting from the selection of the system's composition, via upgrade strategies and system extensions to the connection and the integration with other systems, potential limitations are something that must be considered. Being challenged with the issues outside of our control results in an environment wherein every step we ought to pursue the highest possible effectiveness in every single activity related to any chunk of our software.

In this paper, we presented the challenges we had in the development of the MIS system used in the geographic area where hardware availability is an issue. We described our approach to system design and maintenance and pointed out the decisions and potential trade-offs we had to apply to make the system the most possibly effective in the given situation.

ACKNOWLEDGMENT

This work has been partially supported by the Ministry of Education, Science, and Technological Development of the Republic of Serbia (grant Nr 451-03-68/2022-14/ 200102)

This work has been partially supported by cost action CERCIRAS (CA19135 - Connecting Education and Research Communities for an Innovative Resource Aware Society)